\documentclass[12pt]{article}
\usepackage{a4}
\usepackage{amssymb}
\usepackage{cite}
\newcommand{\be}{\begin{equation}}
\newcommand{\ee}{\end{equation}}
\newcommand{\bea}{\begin{eqnarray}}
\newcommand{\eea}{\end{eqnarray}}

\begin{document}
\thispagestyle{empty}
\def\thefootnote{\fnsymbol{footnote}}
\begin{center}\Large
Defects, Non-abelian T-duality, and the Fourier-Mukai transform of the Ramond-Ramond fields.\\
\vskip 2em
October 2013
\end{center}\vskip 0.2cm
\begin{center}
Eva Gevorgyan$^{1}$ 
\footnote{yeva921228@gmail.com}
 and Gor Sarkissian$^{1,2}$\footnote{ gor.sarkissian@ysu.am}
\end{center}
\vskip 0.2cm

\begin{center}
$^1$Department of Theoretical Physics, \ Yerevan State University,\\
Alex Manoogian 1, 0025\, Yerevan\\
Armenia
\end{center}
\begin{center}
$^2$The Abdus Salam ICTP, \\
Strada Costiera 11, 34014\, Trieste\\
Italy
\end{center}
\vskip 1.5em
\begin{abstract} \noindent
We construct topological defects generating non-abelian T-duality for isometry groups acting without isotropy.
We find that these defects are given by line bundles on the correspondence space with curvature
which can be considered as  a non-abelian generalization of the curvature
of the Poincar\`{e} bundle.  We show that the defect equations of motion encode the non-abelian T-duality transformation.
The Fourier-Mukai transform of the Ramond-Ramond fields generated
by the gauge invariant flux of these defects is studied. We show that it provides elegant and compact way of computation of the transformation of the Ramond-Ramond fields under the non-abelian T-duality.\\
\end{abstract}

\newpage
\section{Introduction}
It is well known that dualities play very important role in String theory.
Among them T-duality (for review see \cite{Giveon:1994fu}) is the most well studied one and has enormous 
significance  from the point of view applications as well as
from the conceptual  point of view.
Massless spectrum of String Theory contains NS sector fields
$G_{\mu\nu}$, $B_{\mu\nu}$, and the dilaton $\Phi$, 
and the RR fields ${\cal G}_{\mu_1\ldots \mu_p}$. Transformations of the NS sector fields were found by Buscher \cite{Buscher:1987qj}. To derive the transformation of the RR fields several approaches were developed: via dimensional reduction
\cite{Bergshoeff:1995as,Meessen:1998qm}, vertex operators for RR fields \cite{Polchinski:1996na}, the gravitino supersymmetry transformation \cite{Hassan:1999bv}, pure spinor formalism\cite{Benichou:2008it}.
It was found in \cite{Polchinski:1996na,Hassan:1999bv,Benichou:2008it}, that the transformation rules of the RR fields under T-duality
are encoded in the rotation
of the bispinor field formed by the RR fields:
\be\label{rtr}
\hat{P}=P\Omega^{-1}
\ee
where
$P={e^{\Phi} \over 2}\sum_k{1\over k!}{\cal G}_{\mu_1\ldots \mu_{k}}\Gamma^{\mu_1\ldots \mu_{k}}$ and
$k$ runs the values $k=1,3\ldots 9$ in the case of IIB,
and the values $k=0,2\ldots 10$ in the case of IIA.
The curved indices Gamma matrices are defined as usual by contracting with the tetrads $e^A_{\mu}$. The matrix $\Omega$ is spinor representation 
of the relative twist between the left and right movers.
For example for T-duality in the direction of coordinate 1, it is
spinor representation of the parity operator in the direction 1:  $\Gamma^1\Gamma_{11}$. On the other hand it was suggested in \cite{Hori:1999me}
the topological rule of the RR field transformation, which
sometimes called Fourier-Mukai transform:
\be\label{fmtin}
\hat{\cal G}=\int_{T^n}{\cal G}\wedge e^{\cal F}=\int_{T^n}{\cal G}\wedge e^{\hat{B}-B+\sum_{i=1}^n d\hat{t}_i\wedge dt^i}
\ee
The integral here is the fiberwise integration, 
${\cal G}$ is the sum of gauge invariant RR fields, having even and odd ranks in  IIA and IIB correspondingly,   $t^i$ and 
$\hat{t}_i$ are coordinates on $T^n$ and dual $\hat{T}^n$ correspondingly.
This formula 
is explained in detail in section 4.1.
The formula (\ref{fmtin}) provides very convenient and elegant way to compute
the RR fields transformation. Also as it was noted in \cite{Bouwknegt:2003vb},
the expression (\ref{fmtin}) guarantees that if RR fields ${\cal G}$ satisfy supergravity Bianchi identity, the dual fields $\hat{\cal G}$ satisfy them as well.
One can understand the expression (\ref{fmtin}), remembering \cite{Hori:1997iq}
that T-duality can be elevated to K-theory as Fourier-Mukai transform (which is also related to the  Nahm transform for instantons) having as a kernel the  Poincar\`{e} bundle.
The two-form $\sum_{i=1}^n d\hat{t}_i\wedge dt^i$ appearing in
exponent is curvature of the Poincar\`{e} line bundle.
On other hand it was shown in \cite{Sarkissian:2008dq}
that the two-form ${\cal F}$ in the exponent of (\ref{fmtin}) is gauge invariant flux of topological defect performing T-duality.

Recently in papers \cite{Sfetsos:2010uq,Lozano:2011kb,Itsios:2012dc,Itsios:2013wd,Jeong:2013jfc}, 
RR fields transformation under non-abelian T-duality
\cite{delaOssa:1992vc,Giveon:1993ai,Elitzur:1994ri,
Alvarez:1994np,Lozano:1995jx,Lozano:1996sc,Borlaf:1996na} was studied.
To derive dual forms in these works  were used the formula (\ref{rtr}),
where now $\Omega$ is determined by the relative
twist of left and right movers under the non-abelian T-duality,
and the dimensional reduction.

In this paper we take the route of the generalization of the Fourier-Mukai transform
to the  non-abelian T-duality for isometry groups acting without isotropy. To reach this aim we use the following strategy. As we mentioned, the two-form ${\cal F}$, appearing in the exponent
of the Fourier-Mukai transform,  is a gauge invariant flux of the defect
performing the corresponding transformations. Hence at the beginning we construct defect performing non-abelian T-duality.
Then we use the derived in this way flux to calculate the RR
fields transformation. The key result of this paper is that the non-abelian
Fourier-Mukai transform  of the RR fields reads:
\be
\widehat{\cal G}=\int_G {\cal G}\wedge e^{\hat{B}-B-dx^a\wedge L^a-{1\over 2}x^af_{bc}^aL^b\wedge L^c}
\ee
Here $L^a$ and $f_{bc}^a$ are Maurer-Cartan forms and structure constants of the isometry group respectively,  $x^a$ are dual coordinates.

Paper is organized in the following way.

In section 2 we review non-abelian T-duality. In particular we recall
the duality relations and demonstrate general formulas for the  case of $SU(2)$ principal chiral model. In section 3 we review actions with defects, present defect performing non-abelian T-duality, and show
that the defect equations of motion reproduce the duality relations
derived in section 1. In section 4.1 we review the Fourier-Mukai transform
formula, and recall how it works for abelian T-duality. In section 4.2
using the flux of non-abelian T-duality defect derived in section 3, we derive the Fourier-Mukai transform formula for non-abelian T-duality,
and compute the RR fields transformation for $SU(2)$ isometry group. We obtain that our results in agreement with that of  \cite{Sfetsos:2010uq,Itsios:2012dc}.

\section{Non-abelian T-duality}
Here we recall and collect some facts on non-abelian T-duality
for isometry groups acting without isotropy \cite{Giveon:1993ai,Itsios:2013wd}.
Suppose we have a target space with an isometry group $G$, with generators $T^a$, structure constants $f_{bc}^a$, and coordinates $\theta^a$, and in some coordinates the metric and the  NS two-form
take the form 
\be\label{mtr1}
ds^2=G_{\mu\nu}(Y)d Y^{\mu}dY^{\nu}+
2G_{\mu a}(Y)\Omega^a_kd Y^{\mu}d \theta^{k}
+G_{ab }(Y)\Omega^a_m\Omega^b_kd\theta^{m}d
\theta^{k}
\ee
\be\label{btr1}
B={1\over 2}B_{\mu\nu}(Y)d Y^{\mu}\wedge dY^{\nu}+
B_{\mu a}(Y)\Omega^a_kd Y^{\mu}\wedge d \theta^{k}
+{1\over 2}B_{ab }(Y)\Omega^a_m\Omega^b_k d\theta^{m}\wedge
d\theta^{k}
\ee
where $\Omega^a_k$ are components of the  right-invariant Maurer-Cartan forms $L^a$:
\bea
dg g^{-1}=L^aT_a=\Omega^a_k  d\theta^k T_a
\eea
The background fields depend  on group coordinates $\theta^a$ only through the  Maurer-Cartan forms.
Also as it is clear from the notations they can depend on some spectator coordinates $Y$.
Since  $d(dgg^{-1})=dgg^{-1}\wedge dgg^{-1}$, $L^a$ and $\Omega^a_k$ satisfy the Maurer-Cartan relations
\be\label{mc1}
dL^a={1\over 2}f^a_{bc}L^bL^c
\ee
and 
\be\label{mc2}
\partial_i\Omega^c_j-\partial_j\Omega^c_i=f^c_{ab}
\Omega^a_i\Omega^b_j
\ee

The corresponding Lagrangian density is 

\bea\label{lag1}
L=Q_{\mu\nu}\partial Y^{\mu}\bar{\partial} Y^{\nu}+
Q_{\mu a}\Omega^a_k\partial Y^{\mu}\bar{\partial} \theta^{k}+
Q_{a\mu }\Omega^a_k\partial \theta^{k}\bar{\partial} Y^{\mu}+
Q_{ab }\Omega^a_m\Omega^b_k\partial \theta^{m}\bar{\partial} 
\theta^{k}
\eea
where
\be
Q_{\mu\nu}=G_{\mu\nu}+B_{\mu\nu}\, ,\hspace{0.1cm}
Q_{\mu a}=G_{\mu a}+B_{\mu a}\, ,\hspace{0.1cm}
Q_{a\mu}=G_{a\mu}+B_{a\mu}\, ,\hspace{0.1cm}
Q_{ab}=G_{ab}+B_{ab}\, .
\ee
To find the dual action one can use the Buscher method and write the Lagrangian (\ref{lag1})  in the form 
\bea\label{lag2}
&&L=Q_{\mu\nu}\partial Y^{\mu}\bar{\partial} Y^{\nu}+
Q_{\mu a}\partial Y^{\mu}\bar{A}^{a}+
Q_{a\mu } A^{a}\bar{\partial} Y^{\mu}+
Q_{ab } A^{a}\bar{A} ^{b}\\ \nonumber
&-& x^a(\partial \bar{A}^a-\bar{\partial}A^a-f_{bc}^aA^b\bar{A}^c)
\eea
The equations of motion of the Lagrangian multiplier $x^a$ force
the field strength $F^a_{+-}=\partial \bar{A}^a-\bar{\partial}A^a-f_{bc}^aA^b\bar{A}^c$ to vanish. The solution to these constraints 
is 
\be\label{agg}
A^a=\Omega^a_k\partial \theta^{k}\;\;\; {\rm and}\;\;\; \bar{A}^a=\Omega^a_k\bar{\partial} \theta^{k}.
\ee
Putting this solution into (\ref{lag2}) yields the original action (\ref{lag1}).
On the other hand integrating out gauge fields $A^a$ one obtains: 
\bea\label{ag1}
&&M_{ba}^{-1}(Q_{\mu b}\partial Y^{\mu}+\partial x^b) =-A^a\\ \nonumber
&&M_{ab}^{-1}(\bar{\partial} x^b-Q_{b\mu }\bar{\partial} Y^{\mu})=\bar{A}^a
\eea
where
\be
M_{ab}=Q_{ab}+x^cf_{ab}^c
\ee
Inserting expressions (\ref{ag1})  back in (\ref{lag2}) we find the dual action:
\bea\label{lagerfeld}
\hat{L}=\hat{E}_{\mu\nu}\partial Y^{\mu}\bar{\partial} Y^{\nu}+
\hat{E}_{\mu a}\partial Y^{\mu}\bar{\partial} x^{a}+
\hat{E}_{a\mu }\partial x^{a}\bar{\partial} Y^{\mu}+
\hat{E}_{ab }\partial x^{a}\bar{\partial} 
x^{b}
\eea
where
\bea\label{bag}
&&\hat{E}_{\mu\nu}=Q_{\mu\nu}-Q_{\mu a}M^{-1}_{ab}Q_{b\nu}\\ \nonumber
&&\hat{E}_{\mu a}=Q_{\mu b}M^{-1}_{ba}\\ \nonumber
&&\hat{E}_{a\mu }=-Q_{b\mu }M^{-1}_{ab}\\ \nonumber
&&\hat{E}_{ab}=M^{-1}_{ab}
\eea

Equating (\ref{agg}) and (\ref{ag1}),  one gets the duality relations for non-abelian T-duality \cite{Borlaf:1996na,Lozano:1996sc}
\be\label{tdna1}
M_{ba}^{-1}(Q_{\mu b}\partial Y^{\mu}+\partial x^b) =-\Omega^a_k\partial\theta^k
\ee
\be\label{tdna2}
M_{ab}^{-1}(\bar{\partial} x^b-Q_{b\mu }\bar{\partial} Y^{\mu})=
\Omega^a_k\bar{\partial}\theta^k
\ee
Separating in (\ref{bag}) symmetric and antisymmetric parts we derive metric and NS form of the dual theory:
\be\label{gmm}
\hat{G}_{\mu\nu}=G_{\mu\nu}-{1\over 2}(Q_{\mu a}M^{-1}_{ab}Q_{b\nu}+Q_{\nu a}M^{-1}_{ab}Q_{b\mu})
\ee
\be
\hat{G}_{\mu a}={1\over 2}(Q_{\mu b}M^{-1}_{ba}-Q_{b\mu }M^{-1}_{ab})
\ee

\be
\hat{G}_{ab}={1\over 2}(M^{-1}_{ab}+M^{-1}_{ba})
\ee

\be
\hat{B}_{\mu\nu}=B_{\mu\nu}-{1\over 2}(Q_{\mu a}M^{-1}_{ab}Q_{b\nu}-Q_{\nu a}M^{-1}_{ab}Q_{b\mu})
\ee
\be
\hat{B}_{\mu a}={1\over 2}(Q_{\mu b}M^{-1}_{ba}+Q_{b\mu }M^{-1}_{ab})
\ee

\be\label{baa}
\hat{B}_{ab}={1\over 2}(M^{-1}_{ab}-M^{-1}_{ba})
\ee
Let us recall  the $SU(2)$ Principal Chiral Model \cite{Fridling:1983ha,Fradkin:1984ai,Sfetsos:2010uq}
\be
S(g)=\int k{\rm Tr}(g^{-1}\partial g g^{-1}\bar{\partial} g)
\ee
where $g\in SU(2)$. The metric in the Euler coordinates is
\be
ds^2=k(d\theta^2+d\phi^2+d\psi^2+2\cos\theta d\phi d\psi)
\ee
and there is no NS two-form.  To obtain the dual background one
should compute $M^{-1}_{ab}$ matrix.

Denoting the dual coordinates $x^a$, $a=1,2,3$,  one has here
\be
M_{ab}=k\delta_{ab}+\epsilon_{abc}x_c
\ee
and
\be
M^{-1}_{ab}={1\over k^2+r^2}\left(k\delta_{ab}+{x_ax_b\over k}
-\epsilon_{abc}x_c\right)
\ee
Separating symmetric and antisymmetric parts and denoting $r^2=x^ax^a$ one gets
\be
\hat{G}_{ab}={1\over k^2+r^2}\left(k\delta_{ab}+{x_ax_b\over k}\right)
\ee
\be
\hat{B}_{ab}=-{1\over k^2+r^2}\epsilon_{abc}x_c
\ee
\be
\hat{\Phi}=-{1\over 2}\log(k^3+kr^2)
\ee
and hence one has
\be\label{mf}
\hat{ds}^2={dr^2\over k}+{kr^2\over k^2+r^2}ds^2(S^2)
\ee
\be\label{bf}
\hat{B}=-{r^3\over k^2+r^2}{\rm Vol}(S^2)
\ee
\section{Non-abelian T-duality via defects}
\subsection{Sigma model with defect}
Defects in two-dimensional quantum field theory are lines
separating different quantum field theories.

Conformal defects are required to satisfy \cite{Bachas:2001vj}

\be
T^{(1)}-\bar{T}^{(1)}=T^{(2)}-\bar{T}^{(2)} \label{ConfDef}
\ee

Topological defects satisfy \cite{Petkova:2000ip}
\be
T^{(1)}=T^{(2)}\, ,\hspace{1cm} \bar{T}^{(1)}=\bar{T}^{(2)} \label{TopDef}
\ee
Since the stress-energy tensor is a generator of diffeomorphisms, condition (\ref{TopDef})
implies that the defect is invariant under a deformation of the line to which it is
attached. A fusion between a defect and a boundary is defined in the case
of topological defects, since the defect can be moved to the boundary without changing the correlator \cite{Petkova:2001zn}.

Let us  briefly review the construction of an action with defects \cite{Fuchs:2007fw,Sarkissian:2008dq}. Let us locate the defect at the vertical line $Z$ defined by the condition $\sigma=0$.
Denote by $\Sigma_1$ the left half-plane $(\sigma\leq 0)$, and by $\Sigma_2$ the right half-plane $(\sigma\geq 0)$, and
a pair of maps $X: \Sigma_1\rightarrow M_1$ and $\tilde{X}: \Sigma_2   \rightarrow M_2$, where $M_1$ and $M_2$ are the target spaces for the two theories.
Assume we have a submanifold $Q$ of the product of target spaces: $Q\subset M_1\times M_2$, with a
connection one-form $A$, and a combined map :
\bea
\Phi : Z\rightarrow M_1\times M_2\\ \nonumber
z\mapsto (X(z), \tilde{X}(z))
\eea
which takes values  in the submanifold $Q$. 

 In this setup one can write the action:
\be\label{defact}
S=\int_{\Sigma_1}L_1dx^+dx^-+\int_{\Sigma_2}L_2dx^+dx^-+\int_Z \Phi^*A
\ee
where
\be
L_i=E^{(i)}_{mn}\partial X^m\bar{\partial}X^n\, ,\;\;\;\;\;\;
E^{(i)}=G^{(i)}+B^{(i)},\;\;\; i=1,2,
\ee
and
\be\label{lccor}
x^{\pm}=\tau\pm \sigma\, .
\ee

\subsection{Defects implementing non-abelian T-duality}
Consider the action  (\ref{defact}) with a defect as in the situation above, where $M_1$ is the target space with the coordinates $(Y^{\mu}, \theta^k)$ and has metric and NS 2-form given by (\ref{mtr1}) and (\ref{btr1}), $M_2$ is the space with the coordinates $(Y^{\mu},x^a)$ and with metric and 2-form given by  (\ref{gmm})-(\ref{baa}),
and $Q$ is the correspondence space, with the  coordinates $(Y^{\mu},\theta^k,x^a)$, the connection
\be\label{conik}
A=-x^aL^a=-x^a\Omega^a_k d\theta^k
\ee
and the curvature
\be\label{curvon}
F=dA=-(dx^aL^a+{1\over 2}x^af_{bc}^aL^bL^c)
\ee
To derive (\ref{curvon}) we used the Maurer-Cartan relation 
(\ref{mc1}).
By other words we take as $L_1$ in (\ref{defact}) the $L$ given by
(\ref{lag1}), and as $L_2$ the $\hat{L}$ given by (\ref{lagerfeld}).

The conditions (\ref{conik}) and (\ref{curvon}) define a line bundle ${\cal P}^{\rm NA}$ over $Q$, with
the curvature (\ref{curvon}), which can be called
non-abelian Poincar\`{e} line bundle.
In this case the action (\ref{defact}) yields the following equations of motion on the defect line:
\bea\label{td11}
&&Q_{\mu a}\partial Y^{\mu}+
Q_{ba}\Omega^b_m\partial \theta^{m}
-Q_{a\mu}\bar{\partial} Y^{\mu}
-Q_{ab }\Omega^b_m \bar{\partial} 
\theta^{m}=\\ \nonumber
&&-x^c\Omega^b_m f_{ba}^c\partial_{\tau}\theta^m
-\partial_{\tau}x^a
\eea

 
\be\label{td33}
\hat{E}_{\mu a}\partial Y^{\mu}+\hat{E}_{b a}\partial x^b
-\hat{E}_{a\mu}\bar{\partial} Y^{\mu}
-\hat{E}_{ab}\bar{\partial} x^{b}=
-\Omega^a_k \partial_{\tau}\theta^k.
\ee
\bea\label{td22}
&&Q_{\mu\alpha}\partial Y^{\mu}+Q_{a\alpha}\Omega^a_k\partial\theta^k
-Q_{\alpha \mu}\bar{\partial} Y^{\mu}-Q_{\alpha a}\Omega^a_k\bar{\partial}\theta^k\\ \nonumber
&&-\hat{E}_{\mu\alpha}\partial Y^{\mu}-
\hat{E}_{a\alpha}\partial x^a
+\hat{E}_{\alpha \mu}\bar{\partial} Y^{\mu}
+\hat{E}_{\alpha a}\bar{\partial} x^a=0 
\eea
In the first line we used the second of the Maurer-Cartan relations (\ref{mc2}).

Solving equations (\ref{td11})-(\ref{td22}) we obtain the duality relations of non-abelian T-duality (\ref{tdna1}) and (\ref{tdna2})

\be\label{tr1}
Q_{\mu a}\partial Y^{\mu}+M_{ba}\Omega^b_m\partial \theta^m=
-\partial x^a
\ee
\be\label{tr2}
Q_{a\mu }\bar{\partial} Y^{\mu}+M_{ab}\Omega^b_m\bar{\partial} \theta^m=
\bar{\partial} x^a
\ee

\be\label{tr3}
M_{ba}^{-1}(Q_{\mu b}\partial Y^{\mu}+\partial x^b) =-\Omega^a_m\partial \theta^m
\ee
\be\label{tr4}
M_{ab}^{-1}(\bar{\partial} x^b-Q_{b\mu }\bar{\partial} Y^{\mu})=\Omega^a_m\bar{\partial} \theta^m
\ee
Using expressions (\ref{bag}) and the duality relations (\ref{tr1}) and  (\ref{tr2}) we obtain
\be
T=\hat{T} \hspace{1cm} {\rm and} \hspace{1cm} \bar{T}=\hat{\bar{T}}
\ee
where
\be
T=G_{\mu\nu}\partial Y^{\mu}\partial Y^{\nu}+
2G_{\mu a}\Omega^a_k\partial Y^{\mu}\partial \theta^{k}
+G_{ab }\Omega^a_m\Omega^b_k\partial\theta^{m}\partial
\theta^{k}
\ee
\be
\bar{T}=G_{\mu\nu}\bar{\partial} Y^{\mu}\bar{\partial} Y^{\nu}+
2G_{\mu a}\Omega^a_k\bar{\partial} Y^{\mu}\bar{\partial} \theta^{k}
+G_{ab }\Omega^a_m\Omega^b_k\bar{\partial}\theta^{m}\bar{\partial}
\theta^{k}
\ee
\be
\hat{T}=\hat{G}_{\mu\nu}\partial Y^{\mu}\partial Y^{\nu}+
2\hat{G}_{\mu a}\partial Y^{\mu}\partial x^{a}
+\hat{G}_{ab }\partial x^{a}\partial
x^{b}
\ee
\be
\hat{\bar{T}}=\hat{G}_{\mu\nu}\bar{\partial} Y^{\mu}\bar{\partial} Y^{\nu}+
2\hat{G}_{\mu a}\bar{\partial} Y^{\mu}\bar{\partial} x^{a}
+\hat{G}_{ab }\bar{\partial}x^{a}\bar{\partial}
x^{b}
\ee
what means that the defect is topological.
\section{Transformation of the Ramond-Ramond fields}
\subsection{Defects and Fourier-Mukai transform}
 As we mentioned, a topological defect can be fused with a boundary, producing new  boundary condition from the old one.
From the other side boundary conditions correspond to D-branes, which can be characterized by their RR charges or 
by elements of the K-theory. Therefore  an action of the defect  on the Ramond-Ramond charges and K-theory elements can be defined.
It is expected \cite{Fuchs:2007fw,Brunner:2008fa,Sarkissian:2008dq,Sarkissian:2010bw,Bachas:2012bj,Elitzur:2013ut} that the action should be ``Fourier-Mukai" type with a kernel given by the exponent of the gauge invariant flux ${\cal F}=\hat{B}-B+F$ on defect,  or by the defect bundle ${\cal P}$ correspondingly.
Saying   Fourier-Mukai type transform we mean the following construction.\footnote{The paragraph below is neither a rigorous nor a precise definition of the Fourier-Mukai transform,
and only has a goal to outline basic ideas. For the rigorous definitions see \cite{bbr,huy} and references therein.} . Suppose we can associate to a  target space $X$ a ring $D(X)$ ( e.g. cohomology groups, K-theory groups, etc.), in a way that for a map $p: X_1\to X_2$ exist pullback $p^*: D(X_2)\to D(X_1)$ and pushforward
$p_*: D(X_1)\to D(X_2)$ maps.
Assume one has an element $K\in D(X\times Y)$. Now we can define the Fourier-Mukai transform, $FM(F)$: $D(X)\to D(Y)$ with the kernel $K$ by the formula:
\be
FM(F)=p^Y_*(K\cdot p^{X*}F)
\ee
where $F\in D(X)$, and  $p^X:X\times Y\to X$, $p^Y:X\times Y\to Y$ are projections.
One can see that usual Fourier transform has this form with the  Riemann integral as pushforward map.

Consider for example the T-duality transformation of the RR fields.

It is found in \cite{Hori:1999me} that  the Ramond-Ramond  fields
of the theory on $T^n\times M$ and those of the T-dual theory on $\hat{T}^n\times M$ are related by a Fourier-Mukai transform:

\be\label{fmt}
\hat{\cal G}=\int_{T^n}{\cal G}\wedge e^{\cal F}=\int_{T^n}{\cal G}\wedge e^{\hat{B}-B+\sum_{i=1}^n d\hat{t}_i\wedge dt^i}
\ee
Here $B$ is the Neveu-Schwarz $B$-field and ${\cal G}=\sum_p {\cal G}_{p}$ is the sum of gauge invariant RR field strength where the sum is over
$p=0,2,4,\ldots$ for Type IIA and $p=1,3,\ldots$ for Type IIB.
The integrand in (\ref{fmt}) is considered as a differential form on the space $M\times T^n\times \hat{T}^n$ and pushforward map is
 fiberwise integration $\int_{T^n}$, mapping differential forms on  $M\times T^n\times \hat{T}^n$ to differential forms  on $M\times \hat{T}^n$. The integral acts on the differential forms of the highest degree  $n$ in $dt_i$ and sets to zero differential forms of lower degree in $dt_i$ \cite{bott}:
\bea\label{fbs}
f(x,\hat{t}_i,t^i)p^*\omega \wedge dt_{i_1}\wedge \ldots dt_{i_r}\mapsto 0, \hspace{1cm} r<n\\ \nonumber
f(x,\hat{t}_i,t^i)p^*\omega \wedge dt_{1}\wedge \ldots dt_{n}\mapsto \omega\int_{T^n}   f(x,\hat{t}_i,t^i)dt_{1}\ldots dt_{n}
\eea
Here $p$ is the projection $M\times T^n\times \hat{T}^n\to M\times \hat{T}^n$,  $\omega$ is a differential form on $M\times \hat{T}^n$, $f(x,\hat{t}_i,t^i)$
is an arbitrary function
and $x$ denotes a point in $M$. 

Since the gauge invariant flux ${\cal F}$ satisfies the condition
\be\label{fhhf}
d{\cal F}=\hat{H}-H
\ee
and the exterior differentiation $d$ commutes with the fiberwise integration \cite{bott},
one can show that the dual forms satisfy the equation \cite{Bouwknegt:2003vb}:
\be\label{fhh}
(d-{\hat H})\wedge\hat{\cal G}=\int_{T^n}e^{\cal F}\wedge(d-H)\wedge{\cal G}
\ee
This implies that $d_H=d-H$ closed forms mapped to 
$d_{\hat H}=d-{\hat H}$ closed form. This means that if the RR fields ${\cal G}$ satisfy  the supergravity
Bianchi identity, so do the dual RR fields $\hat{\cal G}$.

The kernel of the Fourier-Mukai transform (\ref{fmt}) is  indeed the exponent of the gauge invariant combination of the $B$ fields and the flux $\sum_{i=1}^n d\hat{t}_i\wedge dt^i$ of the T-duality defect given by the Poincar\'{e}
bundle \cite{Sarkissian:2008dq} (many detailed explanations
on the T-duality defects can be also found in \cite{Elitzur:2013ut})
\be
e^{\cal F}=e^{\hat{B}-B+\sum_{i=1}^n d\hat{t}_i\wedge dt^i}
\ee
As preparation to the calculations for the case of non-abelian T-duality
in the next section,
now we show  how the formula (\ref{fmt}) produces the known
transformation rules of the Ramond-Ramond fields for the case
of the abelian T-dualization in the direction of one coordinate, which we choose to be the first one.
Remember that in this case the Buscher  transformation rules
of the metric $G$ and NS two-form $B$ are:
\bea\label{tdualc}
&&\hat{G}_{11}={1\over G_{11}}\\ \nonumber
&&\hat{G}_{1M}={B_{1M}\over G_{11}}\\ \nonumber
&&\hat{B}_{1M}={G_{1M}\over G_{11}}\\ \nonumber
&&\hat{G}_{MN}=G_{MN}-{1\over G_{11}} (G_{M1}G_{1N}+B_{1N}B_{M1})\\ \nonumber
&&\hat{B}_{MN}=B_{MN}-{1\over G_{11}} (G_{M1}B_{1N}+G_{1N}B_{M1})
\eea
Here capital latin letters run from 2 to the dimension of the target
spaces.
With (\ref{tdualc})  at hand (\ref{fmt}) takes the form:
\be\label{fmt2}
\hat{\cal G}=\int_{S^1}{\cal G}\wedge e^{(A_1+d\hat{t}^1)\wedge(A_2+dt^1)}=
\int_{S^1}{\cal G}\wedge (1+(A_1+d\hat{t}^1)\wedge (A_2+dt^1))
\ee
where 
\be
A_1=B_{1N}dX^N\:\:\: {\rm and}\:\:\: A_2={G_{1N}\over G_{11}}dX^N
\ee
Taking ${\cal G}$ in the form
\be
{\cal G}={\cal G}^{(0)}+{\cal G}^{(1)}\wedge dt^1
\ee
and using the rules (\ref{fbs}), one obtains
\be
\widehat{\cal G}=\widehat{{\cal G}}^{(0)}+\widehat{{\cal G}}^{(1)}\wedge d\hat{t}^1
\ee
where
\be
\widehat{\cal G}^{(0)}={\cal G}^{(1)}+{\cal G}^{(0)}\wedge A_1+{\cal G}^{(1)}\wedge A_1\wedge A_2
\ee
and
\be
\widehat{\cal G}^{(1)}={\cal G}^{(0)}-{\cal G}^{(1)}\wedge A_2
\ee
These are indeed the RR fields transformation rules
under the abelian T-duality \cite{Meessen:1998qm}.

\subsection{Non-abelian T-duality Fourier-Mukai transform of the Ramond-Ramond fields}
Taking into account that  the  curvature of the defect generating 
the non-abelian T-duality is given by the formula (\ref{curvon}),
the Fourier-Mukai transform of the RR fields takes the form:

\be\label{fmna}
\widehat{\cal G}=\int_G {\cal G} \wedge e^{\hat{B}-B-dx^a\wedge L^a-{1\over 2}x^af_{bc}^aL^b\wedge L^c}
\ee
Here we apply this formula to the case of background 
considered in \cite{Sfetsos:2010uq,Itsios:2012dc}, namely:
\be\label{bgo}
ds^2=ds^2(M_7) +k(Y)ds^2(S^3)
\ee

Here $M_7$ is  a seven-dimensional manifold, $Y$ are  coordinates on $M_7$, $k(Y)$ is a function of $Y$. One can have also $B$ field
on $M_7$.
 The second term is actually
the $SU(2)$ principal chiral model, considered in section 2.
Therefore, using formulae (\ref{mf}) and  (\ref{bf}) the dual model takes the form:
\be
\widehat{ds}^2=ds^2_{M_7}(Y) +{dr^2\over k}+{kr^2\over k^2+r^2}ds^2(S^2)
\ee
and
\be\label{bf4}
\hat{B}=B-{r^3\over k^2+r^2}{\rm Vol}(S^2)
\ee

Consider the following RR forms:
\be
{\cal G}=
{\cal G}^{(0)}+{\cal G}^{(1)}_a\wedge L^a+{1\over 2}{\cal G}^{(2)}_{ab}\wedge L^a\wedge L^b+{\cal G}^{(3)}\wedge L^1\wedge L^2\wedge L^3
\ee
Here ${\cal G}^{(0)}$, ${\cal G}^{(1)}$, ${\cal G}^{(2)}$, ${\cal G}^{(3)}$ are forms on $M_7$.

Denote the forms in the exponent of (\ref{fmna}) as
\be\label{a20}
A^{(2,0)}=\hat{B}-B
\ee
\be
A^{(1,1)}=-dx^a\wedge L^a
\ee
\be
A^{(0,2)}=-{1\over 2}x^af_{bc}^aL^b\wedge L^c
\ee
In this notations we  indicate by the first number the
degree of the form in $dx^a$, and by the second in $L^a$.
Expanding the exponent and remembering that one can have at most
third degree terms in the both kinds of 1-forms we get:
\bea
&&e^{\hat{B}-B-dx^a\wedge L^a-{1\over 2}x^af_{bc}^aL^b\wedge L^c}=1+A^{(2,0)}+A^{(1,1)}+A^{(0,2)}+\\ \nonumber
&&{1\over 2}A^{(1,1)}\wedge A^{(1,1)}+A^{(2,0)}\wedge A^{(1,1)}+
A^{(1,1)}\wedge A^{(0,2)}+A^{(2,0)}\wedge A^{(0,2)}+\\ \nonumber
&&{1\over 6}A^{(1,1)}\wedge A^{(1,1)}\wedge A^{(1,1)}+
A^{(2,0)}\wedge A^{(1,1)}\wedge A^{(0,2)}
\eea
Using the rules of the fiberwise integration we obtain that the dual of the first term comes from the all third order terms in $L^a$ appearing in the expansion of the exponent:
\be
\widehat{{\cal G}^{(0)}}={\cal G}^{(0)}\wedge\omega^{(3)}
\ee
where
\be
\omega^{(3)}=
\int_G {1\over 6}A^{(1,1)}\wedge A^{(1,1)}\wedge A^{(1,1)}+
A^{(2,0)}\wedge A^{(1,1)}\wedge A^{(0,2)}+A^{(0,2)}\wedge A^{(1,1)}
\ee
One can explicitly compute that
\be
{1\over 6}A^{(1,1)}\wedge A^{(1,1)}\wedge A^{(1,1)}
=dx^1\wedge dx^1\wedge dx^2\wedge {\rm vol}(SU(2))
\ee
where we introduced ${\rm vol}(SU(2))=L^1\wedge L^2\wedge L^3$,
\be
A^{(0,2)}\wedge A^{(1,1)}=x^adx^a\wedge {\rm vol}(SU(2))=rdr\wedge {\rm vol}(SU(2))
\ee
\be\label{rdrl}
A^{(2,0)}\wedge A^{(1,1)}\wedge A^{(0,2)}=-{r^4dr\over k^2+r^2}\wedge{\rm Vol}(S^2)
\wedge {\rm vol}(SU(2))
\ee
To derive (\ref{rdrl}) we used the expressions (\ref{bf4}) and (\ref{a20}) for $A^{(2,0)}$.
Collecting all and using that $dx^1\wedge dx^2\wedge dx^3=r^2dr\wedge{\rm vol}(S^2)$
we obtain
\be
\widehat{{\cal G}^{(0)}}={\cal G}^{(0)}\wedge \left({r^2k^2dr\over k^2+r^2}\wedge {\rm vol}(S^2)+rdr\right)
\ee

Similarly collecting all the second order terms  in $L^a$ in the expansion of the exponent one obtains the dual of the second term:
\bea
&&\widehat{{\cal G}^{(1)}_a\wedge L^a}=\int_G {1\over 2}{\cal G}^{(1)}_a\wedge L^a \wedge A^{(1,1)}\wedge A^{(1,1)}+
 {\cal G}^{(1)}_a\wedge L^a \wedge A^{(0,2)}+\\ \nonumber
&&\int_G {\cal G}^{(1)}_a\wedge L^a \wedge A^{(2,0)} \wedge A^{(0,2)}
 =-{1\over 2}\epsilon_{abc}{\cal G}^{(1)}_a\wedge dx^b\wedge dx^c-{\cal G}^{(1)}_ax^a-A^{(2,0)}\wedge{\cal G}^{(1)}_ax^a
\eea
Picking up the first order terms  in $L^a$ gives us the dual of the third term:
\bea
\widehat{{\cal G}^{(2)}_{ab}\wedge L^a\wedge L^b}=\int_G {\cal G}^{(2)}_{ab}\wedge L^a\wedge L^b\wedge  A^{(1,1)}
+{\cal G}^{(2)}_{ab} \wedge L^a\wedge L^b\wedge A^{(1,1)}\wedge A^{(2,0)}\\ \nonumber
=-\epsilon_{abc}{\cal G}^{(2)}_{ab} \wedge dx^c+\epsilon_{abc}{\cal G}^{(2)}_{ab}x^c \wedge{r^2dr\over k^2+r^2}\wedge {\rm vol}(S^2)
\eea
And finally the dual of the last term is given by the terms 
not containing $L^a$ at all:
\be
\int_G {\cal G}^{(3)}\wedge L^1\wedge L^2\wedge L^3 \wedge e^{\hat{B}-B-dx^a\wedge L^a-{1\over 2}x^af_{bc}^aL^b\wedge L^c}={\cal G}^{(3)}+{\cal G}^{(3)}\wedge (\hat{B}-B)
\ee
Rearranging the terms in order of $dx^a$ we can write for the non-abelian T-dual of  ${\cal G}$:
\be
\widehat{{\cal G}}=\widehat{{\cal G}^{(0)}}+\widehat{{\cal G}^{(1)}}+\widehat{{\cal G}^{(2)}}+\widehat{{\cal G}^{(3)}}
\ee
where
\be\label{imr1}
\widehat{{\cal G}^{(0)}}=-{\cal G}^{(1)}_ax^a+{\cal G}^{(3)}
\ee
\be\label{imr2}
\widehat{{\cal G}^{(1)}}={\cal G}^{(0)}\wedge rdr-{1\over 2}\epsilon_{abc}{\cal G}^{(2)}_{ab}\wedge dx^c
\ee
\be\label{imr3}
\widehat{{\cal G}^{(2)}}=-{1\over 2}\epsilon_{abc}{\cal G}^{(1)}_a\wedge dx^b\wedge dx^c-(\hat{B}-B)\wedge {\cal G}^{(1)}_ax^a+{\cal G}^{(3)}\wedge (\hat{B}-B)
\ee
\be\label{imr4}
\widehat{{\cal G}^{(3)}}={\cal G}^{(0)}\wedge {r^2k^2dr\over k^2+r^2}\wedge {\rm Vol}(S^2)+{1\over 2}\epsilon_{abc}{\cal G}^{(2)}_{ab}x^c\wedge{r^2dr\over k^2+r^2}\wedge {\rm vol}(S^2)
\ee

As we have explained before, since the  gauge invariant flux on the defect, which appears in the exponent of (\ref{fmna}), satisfies the 
relation (\ref{fhhf}) , and the exterior differentiation  commutes with the fiberwise integration, the dual fields satisfy the relation:
\be\label{fmnad}
(d-\hat{H})\wedge\widehat{\cal G}=\int_G  e^{\hat{B}-B-dx^a\wedge L^a-{1\over 2}x^af_{bc}^aL^b\wedge L^c}
\wedge (d-H)\wedge{\cal G}
\ee

The relation (\ref{fmnad})  guarantees
that the hatted forms satisfy the supergravity Bianchi identity
given that  so do the original forms ${\cal G}$.
In  \cite{Sfetsos:2010uq,Itsios:2012dc}, the non-abelian T-duality
transformation of the RR fields was performed
for backgrounds (\ref{bgo}), using the approaches based on equation (\ref{rtr}) and the dimensional reduction,
with the RR fields having the form:
 \be
{\cal G}=
{\cal G}^{(0)}+{\cal G}^{(3)}\wedge L^1\wedge L^2 \wedge L^3
\ee
The results obtained  in these works are in  agreement with the formulae
(\ref{imr1})-(\ref{imr4}) for this case.

\section{Discussion}
One of the exciting direction of the further work is study of the elevation of
the non-abelian T-duality Fourier-Mukai transform to K-theory. As we know the D-branes
are elements of K-theory. The arguments of section 4.1 imply that 
if a brane given by an element D of the K-theory of the space $M_1$, under non-abelian T-duality it is mapped to the following element 
of the K-theory of the dual space $M_2$:
\be
FM(D)=p_{2!}\left({\cal P}^{NA}\otimes p_1^{!}D\right)
\ee
Here $p_1$ and $p_2$ are projections of the correspondence 
space $Q$ to $M_1$ and $M_2$ correspondingly, upper and lower
shrieks  denote pullback and pushforward maps in K-theory
and ${\cal P}^{NA}$ is line bundle on $Q$ with the curvature $F=dx^aL_a+
{1 \over 2}x^af^a_{bc}L^bL^c$ constructed in section 3.2.

Next it is interesting to use the technique developed in this paper  to study Ramond-Ramond fields transformation under non-abelian T-duality for other groups than $SU(2)$.

Another important direction is to generalize these results to
non-abelian T-duality with isometry group acting with isotropy.

\section*{Acknowledgments}
This work was partially supported by ANSEF hepth-3267 grant.

The work of G.S. was supported by grant of Armenian State Council of Science 13-1C278.

G.S. would like also to thank The Racah Institute of Physics, Jerusalem,
Israel, where this work was started, and  especially  Shmuel Elitzur 
for support and valuable discussions.\\

\vspace*{3pt}

\end{document}